\documentclass[twocolumn,showpacs,amsmath,amssymb,aps,prx,footinbib,floatfix,superscriptaddress,longbibliography]{revtex4-1}
\usepackage{graphicx}
\usepackage{braket}
\usepackage{savesym}
\usepackage{amsfonts}
\usepackage{bm}
\usepackage{color}
\usepackage{subfigure}
\usepackage{wasysym}
\usepackage{upgreek}
\usepackage{float}
\usepackage{dsfont}
\usepackage{mathtools}
\usepackage{multirow}
\usepackage[usenames,dvipsnames,svgnames,table]{xcolor}
\usepackage{hyperref}
\usepackage{hhline} 
\usepackage{multirow} 
\usepackage[normalem]{ulem}
\usepackage{soul}

\usepackage{makecell}
\usepackage[english]{babel}

\hypersetup{
	colorlinks=true,       
	linkcolor=VioletRed,  
    citecolor=JungleGreen,        
    filecolor=Orchid,      
    urlcolor=black           
}
\begin{document}

\title{Investigating Topological Order using Recurrent Neural Networks}

\author{Mohamed Hibat-Allah}
\email{mohamed.hibat.allah@uwaterloo.ca}
\affiliation{Vector Institute, MaRS  Centre,  Toronto,  Ontario,  M5G  1M1,  Canada}
\affiliation{Department of Physics and Astronomy, University of Waterloo, Ontario, N2L 3G1, Canada}
\affiliation{Perimeter Institute for Theoretical Physics, 31 Caroline St N, Waterloo, ON N2L 2Y5, Canada}

\author{Roger G. Melko}
\affiliation{Department of Physics and Astronomy, University of Waterloo, Ontario, N2L 3G1, Canada}
\affiliation{Perimeter Institute for Theoretical Physics, 31 Caroline St N, Waterloo, ON N2L 2Y5, Canada}

\author{Juan Carrasquilla}
\affiliation{Vector Institute, MaRS  Centre,  Toronto,  Ontario,  M5G  1M1,  Canada}
\affiliation{Department of Physics and Astronomy, University of Waterloo, Ontario, N2L 3G1, Canada}
\affiliation{ Department of Physics, University of Toronto, Ontario M5S 1A7, Canada}

\date{\today}


\begin{abstract}

Recurrent neural networks (RNNs), originally developed for natural language processing, hold great promise for accurately describing strongly correlated quantum many-body systems. Here, we employ two-dimensional (2D) RNNs to investigate two prototypical quantum many-body Hamiltonians exhibiting topological order. Specifically, we demonstrate that RNN wave functions can effectively capture the topological order of the toric code and a Bose-Hubbard spin liquid on the kagome lattice by estimating their topological entanglement entropies. We also find that RNNs favor coherent superpositions of minimally-entangled states over minimally-entangled states themselves. Overall, our findings demonstrate that RNN wave functions constitute a powerful tool to study phases of matter beyond Landau's symmetry-breaking paradigm.

\end{abstract}

\maketitle

\section{Introduction}

Landau symmetry breaking theory provides a fundamental description for a wide range of phases of matter and their phase transitions through the use of local order parameters~\cite{sachdev_2011}. Despite the fact that a great deal of our theoretical and experimental investigations of interacting quantum many-body systems have been developed with the aim of studying local order parameters, it is well-known that the most intriguing strongly correlated phases of matter may not be easily characterized through these observables. 
Instead, several states of matter seen in modern theoretical and experimental studies are defined globally through the phases' topological properties~\cite{Wen_2013,RydbergSimulator2021,TEE2021}. Topological order, in particular, refers to a type of order characterized by the emergence of quasi-particle anyonic excitations, topological invariants, and long-range entanglement, which typically do not appear in traditional forms of order. As a result of these properties, topologically ordered phases have been suggested as an important building block for the development of a protected qubit resistant to perturbations and errors~\cite{Dennis_2002, kitaevAnyonsExactlySolved2006,kitaev2009topological}. Interestingly, such qubits have been devised recently at the experimental level~\cite{Krinner_2022, Sivak2023, Acharya2023}. 
    
While most manifestations of topological order are dynamical in nature---e.g. anyon statistics, ground state degeneracy, and edge excitations~\cite{LevinWen2006}--topological order can also be characterized directly in terms of the ground state wave function and its entanglement. In particular, a probe for topological order is the topological entanglement entropy (TEE)~\cite{KitaevPreskill2006, LevinWen2006}, which offers a characterization of the global entanglement pattern of topological ground states not present in conventionally ordered systems. Notably, the TEE is readily accessible for large classes of topological orders~\cite{LevinWen2006,Hamma2004}, in numerical simulations based on quantum Monte Carlo (QMC)~\cite{TEE2011,TEE2017, Zhao_2022} and density matrix renormalization group (DMRG)~\cite{TEE2012, Jiang2013}, as well as in experimental realizations of topological order based on gate-based quantum computers~\cite{TEE2021}.  
Machine learning (ML) techniques offer an alternative approach study quantum many-body systems and have proved useful for a wide array of tasks including the classification of phases of matter~\cite{Carrasquilla2017, Broecker2017, PhysRevX.7.031038, Miles_2021}, quantum state tomography~\cite{Torlai2018, Carrasquilla2019}, finding ground states of quantum systems~\cite{androsiuk1993,LAGARIS19971,Carleo2017, zi2018, Di_Luo, Pfau_2020, Hermann_2020, choo_fermionicnqs2020, RNNWF, roth2020iterative}, studying open quantum systems~\cite{Vicentini2019,Luo2022}, and simulating quantum circuits~\cite{Jonsson2018,Medvidovic2021, Carrasquilla_2021}, among many others~\cite{dunjkoMachineLearningArtificial2018,RevModPhys.91.045002,Carrasquilla2020,dawidModernApplicationsMachine2022}. In particular, neural networks representations of quantum many-body states have been shown to be able of expressing topological order using, e.g., restricted Boltzmann machines~\cite{Deng_2017,Glasser_2018, chenEquivalenceRestrictedBoltzmann2018,PhysRevB.99.155136}, convolutional neural networks~\cite{Carrasquilla2017} and autoregressive neural networks~\cite{luo2021gauge}. Here we use recurrent neural networks (RNN)~\cite{hochreiter1997long,graves2012supervised,lipton2015} as an ansatz wave function~\cite{RNNWF, roth2020iterative} to investigate topological order in two dimensions (2D) through the estimation of the TEE. We focus on two model Hamiltonians exhibiting topological order, namely Kitaev's toric code~\cite{kitaevAnyonsExactlySolved2006,kitaev2009topological} and a Bose-Hubbard model on the kagome lattice previously shown to host a gapped quantum spin liquid with non-trivial emergent $\mathbb{Z}_2$ gauge symmetry~\cite{PhysRevLett.97.207204,TEE2011,Zhao_2022}. In our study, we use Kitaev-Preskill constructions~\cite{KitaevPreskill2006}, and finite size-scaling analysis of the entanglement entropy to extract the TEE. We find convincing evidence that RNNs are capable of expressing ground states of Hamiltonians displaying topological order. We also find evidence that the RNN wave function is naturally biased toward finding superpositions of minimally entangled states (MESs), as reflected in the calculations of entanglement entropy and Wilson loop operators for the toric code. Overall, our results indicate that RNNs can represent phases of matter beyond the conventional Landau symmetry-breaking paradigm.

\section{2D RNN}
\label{sec:2DRNN}

\begin{figure*}
    \centering
    \includegraphics[width =\linewidth]{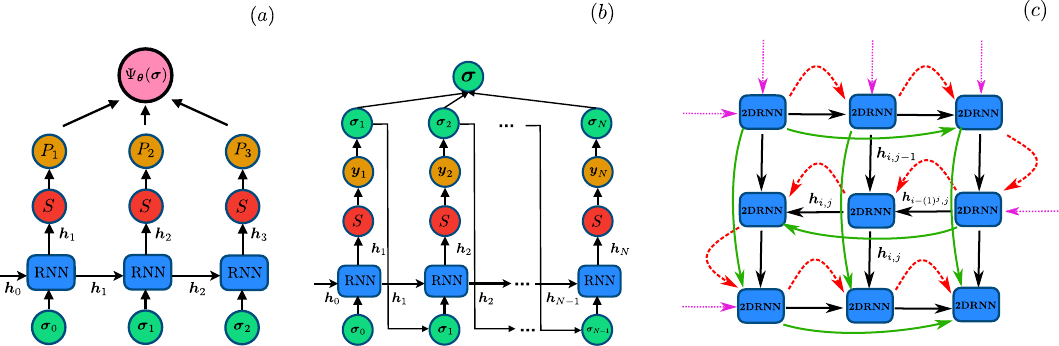}
    \caption{(a) An illustration of the positive RNN wave function. Each RNN cell (blue squares) receives  an input $\bm{\sigma}_{n-1}$ and a hidden state $\bm{h}_{n-1}$ and outputs a new hidden state $\bm{h}_n$. The latter is fed to a Softmax layer (denoted by S, red circle) that outputs a conditional probability $P_i$ (orange circle). The conditional probabilities are multiplied after taking a square root to obtain the wave function $\Psi_{\bm{\theta}}(\boldsymbol{\sigma}) $ (pink circle). (b) Illustration of the sampling scheme for an RNN wave function. After obtaining the probability vector $\bm{y}_i$ from the Softmax layer (S) at step $i$, we sample it to produce $\bm{\sigma}_i$ which is fed again to the RNN with the hidden state $\bm{h}_i$ to produce the next configuration $\bm{\sigma}_{i+1}$.(c) A 2D RNN with periodic boundary conditions. A bulk RNN cell receives two hidden states $\bm{h}_{i,j-1}$ and $\bm{h}_{i-(-1)^j,j}$, as well as two input vectors $\bm{\sigma}_{i,j-1}$ and $\bm{\sigma}_{i-(-1)^j,j}$ (not shown) illustrated by the black solid arrows. To handle periodic boundary conditions, RNN cells at the boundary receive  an additional $\bm{h}_{i,j+1}$ and $\bm{h}_{i+(-1)^j,j}$, as well as two input vectors $\bm{\sigma}_{i,j+1}$ and $\bm{\sigma}_{i+(-1)^j,j}$ (not shown) illustrated by green solid arrows. The sampling path is illustrated with curved dashed red arrows. The initial memory states and the initial inputs are taken as null vectors and are fed to the 2DRNN cells as indicated by the pink dashed arrows.}
    \label{fig:RNN}
\end{figure*}

Our main aim is to study topological properties of Hamiltonians using an RNN wave function ansatz. Since the quantum systems we study are stoquastic~\cite{bravyi2015monte}, we consider an ansatz with positive amplitudes to model the ground state wave function~\cite{RNNWF}. Complex extensions of RNN wave functions for non-stoquastic Hamiltonians have been explored in Refs.~\cite{RNNWF, roth2020iterative}. To model a positive RNN wave function, we write our ansatz in the computational basis as:
\begin{equation*}
   \Psi_{\bm{\theta}}(\bm{\sigma}) = \sqrt{p_{\bm{\theta}}(\bm{\sigma})},
\end{equation*}
where $\bm{\theta}$ denotes the variational parameters of the ansatz $\ket{\Psi_{\bm{\theta}}}$, and $\bm{\sigma} = (\sigma_1, \sigma_2, \ldots, \sigma_N)$ is a basis state configuration. A key characteristic of the RNN wave function is its ability to estimate observables with uncorrelated samples through autoregressive sampling~\cite{RNNWF, Ferris_2012}. This is achieved by parametrizing the joint probability $p_{\bm{\theta}}(\bm{\sigma})$ with its conditionals $p_{\bm{\theta}}(\sigma_i | \sigma_{<i})$, through the probability chain rule
\begin{equation*}
    p_{\bm{\theta}}(\bm{ \sigma})= p_{\bm{\theta}}(\sigma_1)p_{\bm{\theta}}(\sigma_2|\sigma_1) \cdots p_{\bm{\theta}}(\sigma_N|\sigma_{N-1}, \dots, \sigma_2, \sigma_1).
\end{equation*}
The conditionals are given by
\begin{equation*}
    p_{\bm{\theta}}(\sigma_i | \sigma_{<i}) = \bm{y}_i \cdot \bm{\sigma}_i,
\end{equation*}
where $\bm{y}_i = \text{Softmax}(U \bm{h}_{i} + \bm{c})$ and `$\cdot$' is the dot product operation. The hidden state $\bm{h}_{i}$ is calculated recursively as~\cite{lipton2015}
\begin{equation}
    \bm{h}_{i} = f\!(W [\bm{\sigma}_{i-1};\bm{h}_{i-1}] + \bm{b}),
    \label{eq:1DRNN}
\end{equation}
where the input $\bm{\sigma}_{i-1}$ is a one-hot encoding of $\sigma_{i-1}$ and the symbol $[. ; .]$ corresponds to the concatenation of two vectors. Furthermore, $U, W, \bm{b}$, and $\bm{c}$ are learnable weights and biases and $f$ is an activation function. The sequential operation of the RNN is shown in  Fig.~\ref{fig:RNN}(a), where the RNN cell, i.e., the recurrent relation in Eq.~\eqref{eq:1DRNN}, is depicted as a blue square. As each of the conditionals $p_{\bm{\theta}}(\sigma_i | \sigma_{<i})$~\cite{RNNWF} is normalized, the distribution $p_{\bm{\theta}}$, and thus the quantum state $|\Psi_{\bm{\theta}}\rangle$, are normalized. Notably, by virtue of the sequential structure built into the RNN ansatz, it is possible to obtain exact samples from $p_{\bm{\theta}}$ by sampling the conditionals $p_{\bm{\theta}}(\sigma_i | \sigma_{<i})$ sequentially as illustrated in Fig.~\ref{fig:RNN}(b). The sampling scheme is parallelizable and can produce fully uncorrelated samples distributed according to $p_{\bm{\theta}}$ without the use of potentially slow Markov chains~\cite{RNNWF, roth2020iterative}.

As our aim is to study 2D quantum systems with periodic boundary conditions, we use 2D RNNs~\cite{graves2007multidimensional, RNNWF}, through the modification of the 1D relation in Eq.~\eqref{eq:1DRNN} to a recursion that encodes the 2D geometry of the lattice, i.e.,
\begin{align}
    \bm{h}_{i,j} &= f\! \Big(
    W[\bm{\sigma}_{i-(-1)^j,j} ; \bm{\sigma}_{i,j-1}; 
    \bm{h}_{i-(-1)^j,j} ; \bm{h}_{i,j-1}]
    +  \bm{b} \Big).
    \label{eq:2DRNN}
\end{align}
Here $\bm{h}_{i,j}$ is a hidden state with two indices for each site in the 2D lattice, which is computed based on the inputs and the hidden states of the nearest neighboring sites. Note that the $x$-index of the horizontal neighboring site follows the zigzag sampling path illustrated by the red dashed arrows in Fig.~\ref{fig:RNN}(c). This sampling path was motivated in Ref.~\cite{RNNWF}, which allows us to circumvent the use of non-local recurrent connections in our 2D RNN as opposed to other sampling paths. Other alternative orderings are discussed in Ref.~\cite{jain2020locally}. Since $\bm{h}_{i,j}$ contains information about of the history of generated variables $\sigma_{i,j}$, it can be used to compute the conditionals
\begin{equation}
    p_{\bm{\theta}}(\sigma_{i,j}| \sigma_{<i,j}) = \text{Softmax}(U \bm{h}_{i,j} + \bm{c}) \cdot \bm{\sigma}_{i,j}.
    \label{eq:prob_softmax}
\end{equation}
It is worth noting that at the boundaries of the 2D lattice, we use additional inputs to the 2D recursion relation as follows:
\begin{align}
    \bm{h}_{i,j} &= f\! \Big(
    W'[\bm{\sigma}_{i-(-1)^j,j} ; \bm{\sigma}_{i,j-1}; \bm{\sigma}_{i+(-1)^j,j}; \bm{\sigma}_{i,j+1}; \nonumber\\ 
    & \bm{h}_{i-(-1)^j,j} ; \bm{h}_{i,j-1}; \bm{h}_{i+(-1)^j,j}; \bm{h}_{i,j+1}] 
    +  \bm{b}' \Big).
    \label{eq:2DPRNN}
\end{align}
For consistency, $W'$ and $\bm{b}'$ are extensions of the weights $W$ and $\bm{b}$ in Eq.~\eqref{eq:2DRNN}. Additionally, $L_x, L_y$ are respectively the width and the length of the 2D lattice. In our study, we choose $L_x = L_y = L$. The additional variables $\bm{\sigma}_{i+(-1)^j,j}$, $\bm{\sigma}_{i,j+1}$ and hidden states $\bm{h}_{i+(-1)^j,j}$, $\bm{h}_{i,j+1}$ allow modeling systems with periodic boundary conditions such that the ansatz accounts for the correlations between physical degrees of freedom across the boundaries as suggested in Ref.~\cite{luo2021gauge}. For more clarity, periodic boundary conditions on the indices of the hidden states and the inputs are assumed. Furthermore, we note that during the process of autoregressive sampling, if either of the input vectors, in Eq.~\eqref{eq:2DPRNN}, have not been encountered yet, we initialize them to a null vector so we preserve the autoregressive nature of the RNN wave function, as illustrated in Fig.~\ref{fig:RNN}(b). Furthermore, Fig.~\ref{fig:RNN}(c) illustrates the autoregressive sampling path in 2D and how information is being transferred among RNN cells. Importantly, we use an advanced version of 2D RNNs which incorporates a gating mechanism to mitigate the vanishing gradient problem~\cite{RNNWF, Vieijra2021, luo2021gauge, RNNAnnealing}. Additional details can be found in App.~\ref{app:GRU}. We also note that implementing lattice symmetries in our RNN ansatz can be done to improve the variational accuracy as shown in Refs.~\cite{RNNWF,RNNAnnealing}, however, we do not pursue this direction in our study. 

Finally, to highlight the advantages of our RNN approach, we note that the training complexity of one gradient descent step of the 2D RNN wave function is quadratic in the size of hidden states denoted as $d_h$. The latter is very inexpensive compared to projected-entangled pair-states (PEPS), which is $\#$P to contract in general~\cite{Haferkamp_2020} and can be approximately contracted with a scaling $\mathcal{O}(\chi^2\tilde{D}^6)$ 
(where $\tilde{D}$ the PEPS bond dimension and $\chi$ is the bond dimension of the intermediate matrix product state (MPS)~\cite{vanderstraeten_gradient_2016}. It is also worth noting that RNNs have the weight-sharing property which allows the extrapolation of small system size calculations to larger system sizes~\cite{roth2020iterative, RNNAnnealing} as illustrated in Sec.~\ref{sec:BH}. We would also like to point out that the sampling and the inference cost of the RNN is linear in the system size which favors RNNs compared to other autoregressive models in terms of computational cost~\cite{RNNWF, luo2021gauge}.

\subsection{Supplementing RNNs optimization with annealing}
\label{sec:annealing}
To train the parameters of the RNN, we minimize the energy expectation value $E_{\bm{\theta}} = \bra{\Psi_{\bm{\theta}}} \hat{H}\ket{\Psi_{\bm{\theta}}}$ using Variational Monte Carlo (VMC)~\cite{becca_sorella_2017}, where $\hat{H}$ is a Hamiltonian of interest. In the presence of local minima in the optimization landscape of $E_{\bm{\theta}}$, the VMC optimization may get stuck in a poor local optimum~\cite{VNA2021, Bukov_2021}. To ameliorate this limitation, we supplement the VMC scheme with a pseudo-entropy whose objective is to help the optimization escape local minima~\cite{roth2020iterative, VNA2021, RNNAnnealing, Roth2022, Khandoker_2023}. The new objective function is defined as
\begin{equation}
    F_{\bm{\theta}}(n) = E_{\bm{\theta}} - T(n) S_{\rm classical} ( p_{\bm{\theta}} ),
    \label{eq:FreeEnergy}
\end{equation}
where $F_{\bm{\theta}}$ is a variational pseudo free energy. The Shannon entropy $S_{\rm classical}$ of $p_{\boldsymbol{\theta}}\left(\boldsymbol{\sigma}\right)$ is given by
\begin{equation}
    S_{\rm classical} (p_{\bm{\theta}}) = - \sum_{\bm{\sigma}} p_{\bm{\theta}}(\bm{\sigma}) \ln\left(p_{\bm{\theta}}(\bm{\sigma})\right),
    \label{eq:vnEntropy}
\end{equation}
where the sum goes over all possible configurations $\{\bm{\sigma}\}$ in the computational basis. The pseudo-entropy $S_{\rm classical}$ and its gradients are evaluated by sampling the RNN wave function. Furthermore, $T(n)$ is a pseudo-temperature that is annealed from some initial value $T_0$ to zero as follows: $T(n) = T_0 (1-n/N_{\rm annealing})$ where $n \in [ 0, N_{\rm annealing} ]$ and $N_{\rm annealing}$ is the total number of annealing steps. This scheme is inspired by the regularized variational quantum annealing scheme in Refs.~\cite{roth2020iterative, VNA2021, RNNAnnealing}. More details about our training scheme are given in App.~\ref{app:VMC}. We also provide the hyperparameters in App.~\ref{app:hyperparams}.

\subsection{Topological entanglement entropy}
\label{sec:TEE}

A powerful tool to probe topologically ordered states of matter is through the TEE~\cite{Hamma2004,Hamma2005,LevinWen2006,KitaevPreskill2006, Hamma2009, Quasiparticle2012, TEE2011, TEE2017,kimUniversalLowerBound2023}. The TEE can be extracted by computing the entanglement entropy of a spatial bipartition of the system into $A$ and $B$, which together comprise the full system. For many phases of 2D matter, the Renyi-$n$ entropy $S_n(A) \equiv \frac{1}{1-n} \ln(\text{Tr}(\rho^n_A))$ satisfies the area law $S_n(A) = aL_A - \gamma + \mathcal{O}(L^{-1})$. Here $L_A$ is the size of the boundary between $A$ and $B$, $\rho_A=\text{Tr}_B |\Psi \rangle \langle \Psi |$  is the reduced density matrix of subsystem $A$, $|\Psi\rangle$ is the state of the system, and $\gamma$ is the TEE. The latter detects non-local correlations in the ground state wave function and plays the role of an order parameter for topological phases similar to the notion of a local order parameter in phases displaying long-range order. Interestingly, a measure of specific non-zero values of $\gamma$ can be a clear sign of the existence of a topological order in a system of interest. Additionally, since the TEE is shown to be independent of the choice of Renyi index $n$ for a contractible region $A$~\cite{Hamma2009}, we can use the swap trick~\cite{EE2010} with our RNN wave function ansatz~\cite{RNNWF, EE2020} to calculate the second Renyi entropy $S_2$ and extract the TEE $\gamma$. 

To access the TEE $\gamma$, we can approximate the ground state of the system using an RNN wave function ansatz, i.e. $|\Psi_{\boldsymbol{\theta}}\rangle\approx |\Psi\rangle$ for different system sizes followed by a finite-size scaling analysis of the second Renyi entropy. We can also make use of a TEE construction, e.g., the Kitaev-Preskill construction~\cite{KitaevPreskill2006}.
\begin{figure*}
    \centering
    \includegraphics[width =\linewidth]{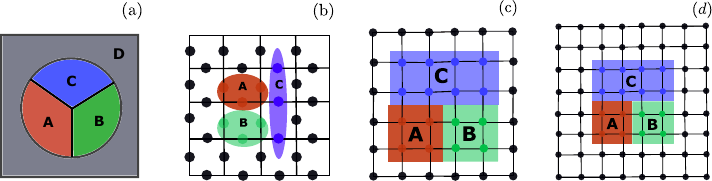}
    \caption{Panel (a): a sketch of the parts $A$, $B$, and $C$ that we use for Kitaev-Preskill construction to compute the TEE in a system of interest. Panels (b-d): an illustration of the subregions $A, B, C$ chosen for the Kitaev-Preskill constructions. (b) For the 2D toric code, each subregion has $3$ spins. We also do the same for the lattice lengths $L = 6,8,10$ in Fig.~\ref{fig:renyi2_tc}(b). For the hard-core Bose-Hubbard model on the Kagome lattice, we target two different system sizes. Panel (c) shows the construction for $L = 6$ and panel (d) provides the construction for $L = 8$. In panels (c) and (d), each site corresponds to a block of three bosons.}
    \label{fig:TEE_constructions}
\end{figure*}

The Kitaev-Preskill construction prescribes dividing the system into four subregions $A$, $B$, $C$, and $D$ as illustrated in Fig.~\ref{fig:TEE_constructions}. The TEE can be then obtained by computing
\begin{align*}
    \gamma &= -S_2(A)-S_2(B)-S_2(C)+S_2(AB)\\
    &+S_2(AC)+S_2(BC) - S_2(ABC),
\end{align*}
where $S_2(A)$ is the second Renyi entropy of the subsystem $A$, and $AB$ is the union of $A$ and $B$ and similarly for the other terms. Finite-size effects on $\gamma$ can be alleviated by increasing the size of the subregions $A, B$ and $C$~\cite{KitaevPreskill2006, Furukawa2007}. Here we have fixed the size of the interior subregions $A$, $B$, $C$ to limit the error
bars in our calculations. Finally, we highlight the ability of the RNN wave function to study systems with fully periodic boundary conditions as a strategy to mitigate boundary effects, as opposed to cylinders used in DMRG~\cite{Stoudenmire2012,Cylinders2014}, which may potentially introduce edge effects that can affect the values of the TEE~\cite{RydbergHarvard2021}. 

\section{Results}
\label{sec:results}

\subsection{The toric code}
\label{sec:TC_results}

\begin{figure}
    \centering
    \includegraphics[width = 0.95\linewidth]{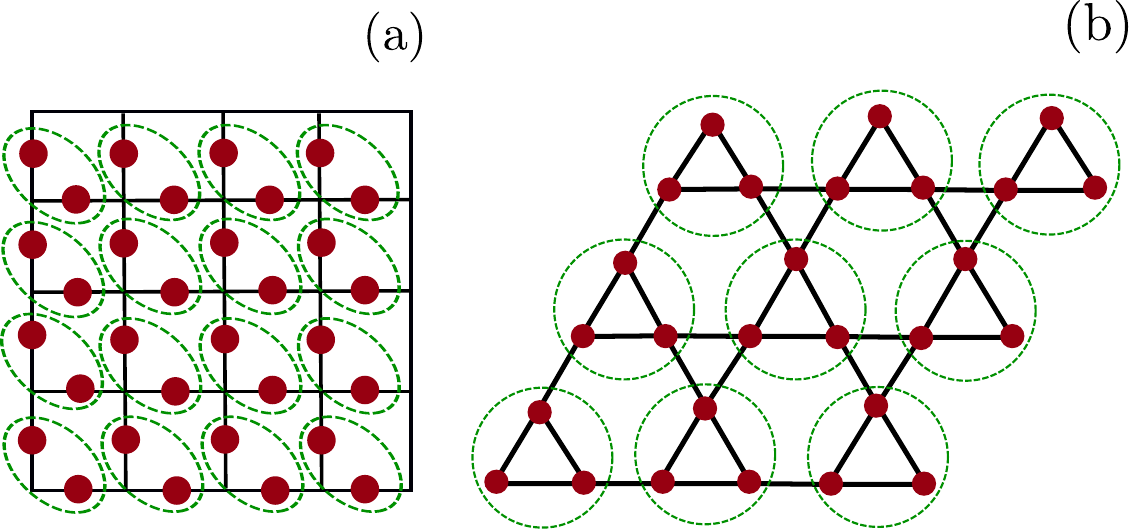}
    \caption{Mapping of 2D toric code lattice and kagome lattice to a square lattice that can be handled by a 2D RNN wave function. For the toric code lattice in panel (a), every two sites inside the dashed green ellipses are merged. For the kagome lattice in panel (c), every three sites in a unit cell enclosed by the dashed green circles are combined.}
    \label{fig:mapping}
\end{figure}

We now focus our attention on the toric code Hamiltonian which is the simplest model that hosts a $Z_2$ topological order~\cite{kitaevAnyonsExactlySolved2006,Hamma2005} and has a non-zero TEE equal to $\gamma = \ln(2)$. The Hamiltonian is defined in terms of spin-$1/2$ degrees of freedom located on the edges of a square lattice (see Fig.~\ref{fig:mapping}(a)) and is given by
\begin{equation*}
    \hat{H} = -\sum_{p} \prod_{i \in p} \hat{\sigma}_i^{z} - \sum_{v} \prod_{i \in v} \hat{\sigma}_i^{x},
\end{equation*}
where $\hat{\sigma}_i^{x,z}$ are Pauli matrices. Additionally, the first summation is on the plaquettes and the second summation is on the vertices of the lattice~\cite{Hamma2005}. Note that the lattice in Fig.~\ref{fig:mapping}(a) can be seen as a square lattice with a unit cell containing two spins. In our simulations, we use an $L \times L \times 2$ array of spins where $L$ is the number of plaquettes on each side of the underlying square lattice. It is possible  to study the toric code with a 2D RNN defined on a primitive square lattice by merging the two spin degrees of freedom of the unit cell of the toric code into a single ``patch" followed by an enlargement of the local Hilbert space dimension in the RNN from $2$ to $4$. This idea is illustrated in Fig.~\ref{fig:mapping}(a) and is similar in spirit to how the local Hilbert space is enlarged in DMRG to study quasi-1D systems~\cite{Milsted2019}. We provide additional details about the mapping in App.~\ref{app:GRU}. 

To extract the TEE from our ansatz, we variationally optimize the 2D RNN wave function targetting the ground state of this model for multiple system sizes on a square lattice with periodic boundary conditions. After the optimization, we compute the TEE using system size extrapolation and using the Kitaev-Preskill scheme provided in Sec.~\ref{sec:TEE}. Here we use three spins for each subregion as illustrated in Fig.~\ref{fig:TEE_constructions}(b). To avoid local minima during the variational optimization, we perform an initial annealing phase as described in Sec.~\ref{sec:annealing} (see additional details in App.~\ref{app:hyperparams}).

\begin{figure}
    \centering
    \includegraphics[width = \linewidth]{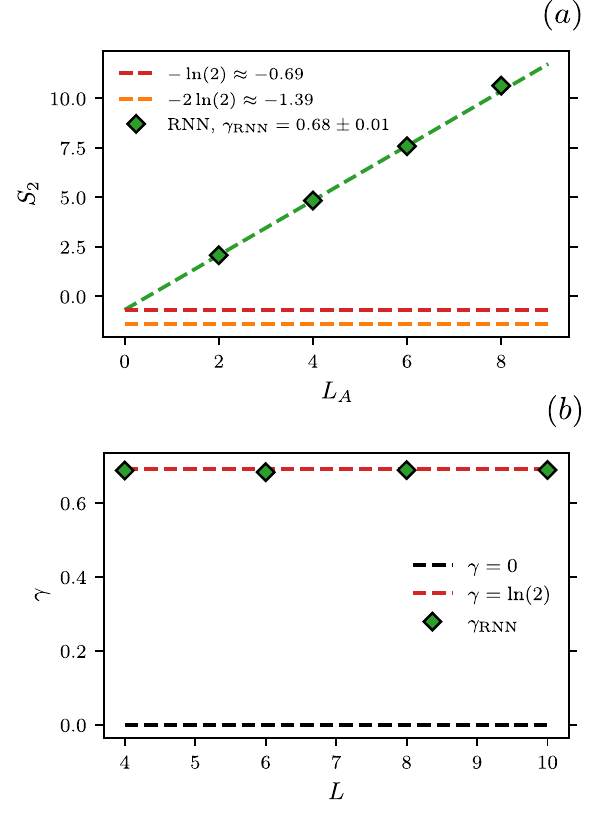}
    \caption{Entanglement properties of the 2D toric code. (a) Second Renyi entropy $S_2$ scaling is computed using our RNN wave function on the 2D toric code for different lengths of the boundary $L_A = L$, where $L$ is the system length and the total system size is given as $L \times L \times 2$. Note that our toric code embedded in a torus is divided into two equal cylinders to compute $S_2$. (b) TEE computed with the Kitaev-Preskill construction (see Fig.~\ref{fig:TEE_constructions}(b)) versus the overall system length $L$. The values found by the RNN are very close to $\ln(2)$. Error bars correspond to one standard deviation and are smaller than the symbol size.}
    \label{fig:renyi2_tc}
\end{figure}

The results shown in Fig.~\ref{fig:renyi2_tc}(a) suggest that our 2D RNN wave function can describe states with an area law scaling in 2D. Linearized versions of the RNN wave function have been recently shown to display an entanglement area law~\cite{Wu2022}. For $L = 10$ (not included in the extrapolations in Fig.~\ref{fig:renyi2_tc}(a)), it is challenging to evaluate $S_2$ accurately as the expectation value of the swap operator is proportional to $\exp{(-S_2)}$, which becomes very small and is hard to resolve accurately via sampling the RNN wave function. The improved ratio trick is an interesting alternative for enhancing the accuracy of our estimates~\cite{EE2010, Torlai2018}. The use of conditional sampling is also another possibility for enhancing the accuracy of our measurements~\cite{EE2020}.

Additionally, the extrapolation confirms the existence of a non-zero TEE whose value is close to $\gamma' = \ln(2)$ within error bars. To further verify that our 2D RNN wave function can extract the correct TEE of the 2D toric code, we compute the TEE using the Preskill-Kitaev construction, which has contractible surfaces, and for which the TEE does not depend on the topological sector superposition~\cite{Quasiparticle2012,TEE2017} (see Fig.~\ref{fig:TEE_constructions}(b)). The results reported in Fig.~\ref{fig:renyi2_tc}(b) demonstrate an excellent agreement between the TEE extracted by our RNN and the expected theoretical value for the toric code. To keep the error bars small, and since in the toric code the TEE does not depend on the subregion sizes~\cite{Hamma2005}, we use fixed subregion sizes in Fig.~\ref{fig:TEE_constructions}(b). 

Interestingly, we note that the subregion we use to compute the TEE in Fig.~\ref{fig:renyi2_tc}(a) is half of the torus, namely a cylinder with two disconnected boundaries~\footnote{Note that this choice allows minimizing the boundary size as opposed to a square region in the bulk. This feature is desirable since the swap operator used to estimate the second Renyi entropy~\cite{RNNWF} becomes very small, and thus more sensitive to statistical errors when the boundary increases for a quantum system satisfying the area law.}. As shown in Ref.~\cite{Quasiparticle2012}, the use of this non-contractible geometry means that the expected TEE becomes state-dependent and given by
\begin{equation}
    \gamma' = 2 \gamma + \ln \left( \sum_i \frac{p_i^2}{d_i^2} \right )
    \label{eq:state_dep_TEE}
\end{equation}
for the second Renyi entropy. Here $d_i \geq 1$ is the quantum dimension of an $i$-th quasi-particle. For the toric code, we have Abelian anyons with $d_i = 1$ for $i = 1,2,3,4$. Additionally $p_i = |\alpha_i|^2$ is the overlap of the computed ground state $\ket{\Psi}$ with the $i$-th MES $\ket{\Xi_i}$ where
\begin{equation*}
    \ket{\Psi} = \sum_{i} \alpha_i \ket{\Xi_i}.
\end{equation*}
The observations above and the numerical result $\gamma'_{\text{RNN}} \approx \ln(2)$, for the non-contractible subregions in Fig.~\ref{fig:renyi2_tc}(a), suggest that the RNN wave functions optimized via gradient descent and annealing find a superposition of MES, as opposed to DMRG which preferentially collapses to a single MES for relatively low bond dimensions. For relatively large bond dimensions a superposition of MES can be recovered in a DMRG simulation~\cite{TEE2012, Jiang2013}. 

\begin{figure}[htp]
    \centering
    \includegraphics[width = 0.7\linewidth]{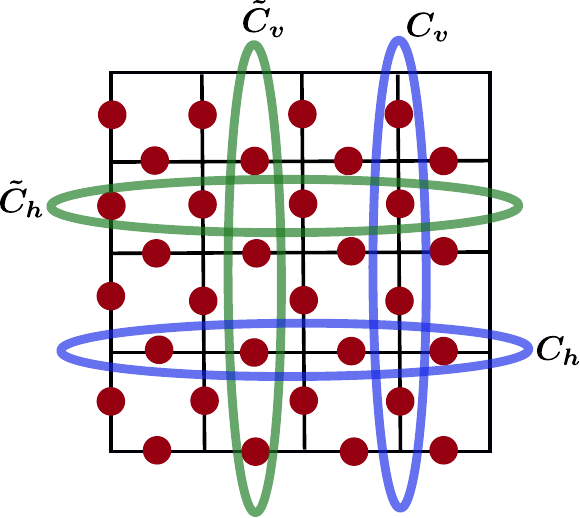}
    \caption{An illustration of the vertical and the horizontal loops used to compute the Wilson loop operators (see Eq.~\ref{eq:Wilson_loop}) and the 't Hooft loop operators (see Eq.~\ref{eq:tHooft_loop}).}
    \label{fig:loop_operators}
\end{figure}

Here we further investigate the superposition found by the RNN through the analyses of the expectation values of the average Wilson loop operators and the average 't Hooft loop operators. First of all, we check that our RNN energy has converged to the ground state energy as shown in Tab.~\ref{tab:ground_state_energies}. This convergence confirms that our RNN wave function satisfies the plaquette and the vertice constraints with an excellent approximation.
\begin{table*}[htb]
\begin{tabular}{|c|c|c|c|c|}
\hline
Length $L$   & $L = 4$     & $L = 6$     & $L = 8$     & $L = 10$    \\ \hline
Energy per spin & -1.99970(3) & -1.99962(2) & -1.99975(1) & -1.99986(1) \\ \hline
\end{tabular}
\caption{Variational energies per site found by our RNN wave function for the 2D toric code for different lattice lengths $L$ after autoregressively sampling $10^5$ configurations at the end of training. We find an excellent agreement with the ground state energy per site $-2$.}
\label{tab:ground_state_energies}
\end{table*}
Next, we define the average Wilson loop operators as 
\begin{equation}
\Hat{W}^{z}_d = \frac{1}{L} \left( \sum_{\mathcal{C}_d} \prod_{\sigma_j \in \mathcal{C}_d} \hat{\sigma}_j^z \right).
\label{eq:Wilson_loop}
\end{equation}
Here $d = h,v$ and $\mathcal{C}_h, \mathcal{C}_v$ are closed non-contractible loops illustrated in Fig.~\ref{fig:loop_operators}. A set of degenerate ground states of the toric code are eigenstates of the operators $\hat{W}^z_h$, $\hat{W}^z_v$ with eigenvalues $\pm 1$. Additionally, the two eigenvalues uniquely determine the topological sector of the ground state. In this case, the topological ground states can be labeled as $\ket{\xi_{ab}}$ with $a,b = 0,1$~\cite{Quasiparticle2012, fradkin_2013}. 

We can also define the average 't Hooft loop operators on non-contractible closed loops~\cite{fradkin_2013}, such that
\begin{equation}
\hat{W}^{x}_d = \frac{1}{L} \left( \sum_{\mathcal{C}_d} \prod_{\sigma_j \in \tilde{\mathcal{C}}_d} \hat{\sigma}_j^x \right),
\label{eq:tHooft_loop}
\end{equation}
where $d = h,v$ and $\tilde{\mathcal{C}}_h$, $\tilde{\mathcal{C}}_v$ correspond to horizontal and vertical loops as illustrated in Fig.~\ref{fig:loop_operators}. These operators satisfy the anti-commutation relations $\{ \hat{W}^z_h, \hat{W}^x_v \} = 0$ and $\{ \hat{W}^z_v, \hat{W}^x_h \} = 0$.

From the optimized RNN wave function ($L = 8$), we find $\langle \hat{W}^z_h \rangle =0.0009(2)$ and $\langle \hat{W}^z_v \rangle =-0.0039(2)$ which are consistent with vanishing expectation values. We also obtain $\langle \hat{W}^x_h \rangle = 0.999847(5)$ and $\langle \hat{W}^x_v \rangle = 0.999785(5)$ for the 't Hooft loop operators, which are consistent with $+1$ expectation values. These results are in part due to the use of a positive RNN wave function which forces the expectation values $\langle \hat{W}^x_h \rangle$ and $\langle \hat{W}^x_v \rangle$ to strictly positive values and rules out the possibility to obtain, e.g., $\langle \hat{W}^x_h \rangle=-1$. 

By expanding the optimized RNN wave function in the $\ket{\xi_{ab}}$ basis, where $a,b$ are binary variables, we obtain
\begin{equation*}
    \ket{\Psi_{\text{RNN}}} \approx \sum_{ab} c_{ab} \ket{\xi_{ab}}.
\end{equation*}
Here $\{\ket{\xi_{ab}} \}$ correspond to the four topological sectors and they are mutually orthogonal. $c_{ab}$ are real numbers since we use a real-valued (specifically positive) ansatz wave function. Additionally, the basis states $\ket{\xi_{ab}}$ satisfy
\begin{align*}
     \hat{W}^z_h \ket{\xi_{ab}} =  (-1)^a \ket{\xi_{ab}}, \\
    \hat{W}^z_v \ket{\xi_{ab}} = (-1)^b \ket{\xi_{ab}}.
\end{align*}
From the anti-commutation relations, we can show that:
\begin{align*}
    \hat{W}^x_h \ket{\xi_{ab}} = \ket{\xi_{a\bar{b}}}, \\
    \hat{W}^x_v \ket{\xi_{ab}} = \ket{\xi_{\bar{a}b}}, \\
\end{align*}
where $\bar{a} = 1-a$ and $\bar{b} = 1-b$. By plugging the last two equations in the $\hat{W}^x_h$, $\hat{W}^x_v$ expectation values of our optimized RNN wave function, we obtain:
\begin{align*}
    2 c_{00}c_{01} + 2 c_{10}c_{11} \approx 1, \\
    2 c_{00}c_{10} + 2 c_{01}c_{11} \approx 1 .
\end{align*}
From the normalization constraint $1 = \sum_{ab} c_{ab}^2$, we deduce that:
\begin{align*}
    (c_{00} - c_{01})^2 + (c_{10} -c_{11})^2 \approx 0, \\
    (c_{00} - c_{10})^2 + (c_{01} - c_{11})^2 \approx 0 .
\end{align*}
As a consequence, we conclude that $c_{00} \approx c_{01} \approx c_{10} \approx c_{11}$, which means that the optimized RNN wave function is approximately a uniform superposition of the four topological ground states $\ket{\xi_{ab}}$. This observation is also consistent with vanishing expectation values of the operators $\hat{W}^z_h$, $\hat{W}^z_v$.

Furthermore, from Ref.~\cite{Quasiparticle2012} the MES of the toric code are given as follows:
\begin{align*}
    \ket{\Xi_1} = \frac{1}{\sqrt{2}} \left( \ket{\xi_{00}} + \ket{\xi_{01}} \right), \\
    \ket{\Xi_2} = \frac{1}{\sqrt{2}} \left( \ket{\xi_{00}} - \ket{\xi_{01}} \right), \\
    \ket{\Xi_3} = \frac{1}{\sqrt{2}} \left( \ket{\xi_{10}} + \ket{\xi_{11}} \right), \\
    \ket{\Xi_4} = \frac{1}{\sqrt{2}} \left( \ket{\xi_{10}} - \ket{\xi_{11}} \right).
\end{align*}
Thus, our RNN wave function can be written approximately as a uniform superposition of the MESs $\ket{\Xi_1}$ and $\ket{\Xi_3}$, i.e.
\begin{equation*}
    \ket{\Psi_{\text{RNN}}} \approx \frac{1}{\sqrt{2}} \left( \ket{\Xi_1} + \ket{\Xi_3} \right).
\end{equation*}
In conclusion, using Eq.~\eqref{eq:state_dep_TEE}, we expect $\gamma' = 2 \ln(2) + \ln\left(\frac{1}{4} + \frac{1}{4} \right) = \ln(2)$, which is consistent with our numerical observations.

We note that the exact autoregressive sampling procedure plays a key role in the ability of our RNN ansatz to sample a superposition of different topological sectors when this superposition is encoded in our ansatz. For wave functions representing the ground state of the toric code used in combination with Markov-chain Monte Carlo methods, the probability of sampling different topological sectors of the state is exponentially suppressed even if the exact wave function ansatz encodes different topological sectors. This observation can be illustrated using an exact convolutional neural network construction of the toric code ground state which contains an equal superposition of different topological sectors~\cite{Carrasquilla2017}. Although in principle such representation contains all topological sectors, its form is not amenable to exact sampling and uses Markov chains so, upon sampling with local moves, the system chooses a fixed topological sector. 

\subsection{Bose-Hubbard model on kagome lattice}
\label{sec:BH}

We now turn our attention to a hard-core Bose-Hubbard model on the kagome lattice, which has been shown to host topological order~\cite{PhysRevLett.97.207204,TEE2011,Zhao_2022}. The Hamiltonian of this model is given by
\begin{equation}
    \hat{H} = -t\sum_{\langle i,j \rangle} \left( b_i^{\dagger} b_j + b_i b_j^{\dagger} \right) + V\sum_{\hexagon} n_{\hexagon}^2,
    \label{eq:bhsl}
\end{equation}
where $b_i$ ($b_i^{\dagger}$) is the annilihation (creation) operator. Furthermore, $t$ is the kinetic strength, $V$ is a tunable interaction strength and $n_{\hexagon} = \sum_{i \in \hexagon} (n_i - 1/2)$. The first term corresponds to a kinetic term that favors hopping between nearest neighbors, whereas the second term promotes an occupation of three hard-core bosons in each hexagon of the kagome lattice. In our setup, we choose $V$ in units of the kinetic term strength $t$. Note that for a hard-core boson $i$, the occupation $n_i = b_i^{\dagger} b_i$ only takes two values $0$ or $1$.

The atom configurations of this model correspond to an $L \times L \times 3$ array of binary degrees of freedom where $L$ is the size of each side of the kagome lattice. Following an analogous approach to the toric code, we combine three sites of the unit cell of the kagome lattice as input to the 2D RNN cell, as illustrated in Fig.~\ref{fig:mapping}(b). This allows us to map our kagome lattice with a local Hilbert space of $2$ to a square lattice with an enlarged Hilbert space of size $2^3 = 8$.

The model is known to host a $Z_2$ spin-liquid phase for $V \gtrsim 7$~\cite{TEE2011,wangTopologicalSpinLiquid2017,Zhao_2022}. To confirm this finding, we estimate $\gamma$ for the system sizes $6 \times 6 \times 3$ and $8 \times 8 \times 3$. We use the Kitaev-Preskill construction~\cite{KitaevPreskill2006}. The details of the construction of the regions $A, B$ and $C$ are provided in Figs.~\ref{fig:TEE_constructions}(c-d). As the Hamiltonian in Eq.~\ref{eq:bhsl} has a $U(1)$ symmetry associated with the conservation of bosons in the system, we impose this symmetry on our RNN wave function~\cite{RNNWF}. We also supplement the VMC optimization with annealing to overcome local minima as previously done for the 2D toric code (see App.~\ref{app:VMC}). For the system size $8 \times 8 \times 3$, the RNN ansatz parameters were initialized using the optimized parameters of the system size $6\times6\times3$ (see details about the hyperparameters in App.~\ref{app:hyperparams}). This pre-training technique was motivated by Refs.~\cite{roth2020iterative, RNNAnnealing, luo2021gauge}.

\begin{figure}
    \centering
    \includegraphics[width =\linewidth]{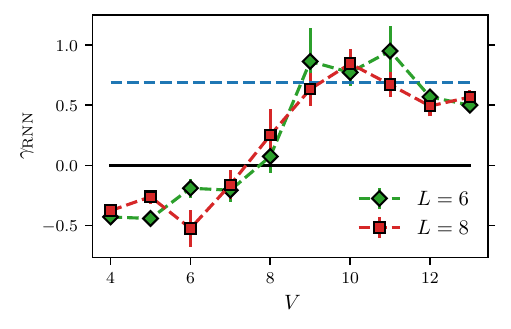}
    \caption{A plot of the topological entanglement entropy against the interaction strength $V$ (in units of $t$) of Hard-core Bose-Hubbard model on kagome lattice for system sizes $N = L \times L \times 3$ where $L = 6,8$. The calculations were performed using the Kitaev-Preskill construction (see Fig.~\ref{fig:TEE_constructions}(c-d)). The continuous black horizontal line corresponds to a zero TEE, and the dashed blue horizontal line for a $\ln(2)$ TEE.}
    \label{fig:BoseHubbard}
\end{figure}

The results are provided in Fig.~\ref{fig:BoseHubbard}. The computed TEEs for $L = 6,8$ show a saturation of $\gamma_{\rm RNN}$ for large values of the interaction strength $V$. We observe that the saturation values of $\gamma_{\rm RNN}$ are in good agreement with the expected TEE $\gamma = \ln(2)$ of a $Z_2$ spin-liquid~\cite{TEE2011}. Additionally, the negative values of $\gamma_{\rm RNN}$ observed for $V \leq 6$ in the superfluid phase~\cite{TEE2011} may be related to the presence of Goldstone modes that manifest themselves as corrections to the area law in the entanglement entropy and can be seen as a negative contribution to the TEE~\cite{Bohdan2015}. We note that the QMC methods are capable of obtaining a consistent value with the exact TEE for this model at $V = 8$ for very large system sizes~\cite{Zhao_2022} using finite-size extrapolation. This observation suggests that our RNN ansatz is still limited by finite-size effects at $V=8$ (see Fig.~\ref{fig:BoseHubbard}) for which the TEE is not yet saturated to $\ln{2}$. Other sources of error in our calculation may be due to inaccuracies in the variational calculations and statistical errors due to the sampling. However, we note that our variational calculation is performed at zero temperature, which makes our calculations insensitive to temperature effects as opposed to QMC~\cite{TEE2011}. 

\section{Conclusions and Outlooks}

We have demonstrated a successful application of neural network wave functions to the task of detecting topological order in quantum systems. In particular, we use RNNs borrowed from natural language processing as ansatz wave functions.
RNNs enjoy the autoregressive property which allows them to sample uncorrelated configurations. They are also capable of estimating second Renyi entropies using the swap trick~\cite{RNNWF} with which we computed TEEs using finite-size scaling and Kitaev-Preskill constructions~\cite{KitaevPreskill2006}. Although we have fixed the size of the interior subregions $A$, $B$, $C$ to limit the error bars in our calculations, we note that it is possible to use improved estimators of TEE~\cite{EE2010, Torlai2018, EE2020}. Furthermore, the structural flexibility of the RNN offers the possibility to handle a wide variety of geometries including periodic boundary conditions in any spatial dimension which alleviate boundary effects on the TEE.

We have empirically demonstrated that 2D RNN wave functions support the 2D area law and can find a non-zero TEE for the toric code and for the hard-core Bose-Hubbard model on the Kagome lattice. We also find that RNNs favor coherent superpositions of MESs over a single MES. The success of our numerical experiments hinges on the combination of the exact sampling strategy used to compute observables, the structural properties of the RNN wave function, and the use of annealing as a strategy to overcome local minima during the optimization procedure. 

The accuracy improvement of our findings can be achieved through the use of more advanced versions of RNNs and autoregressive models in general~\cite{RNNAnnealing, Wu2022}, or even a hybrid approach that combines QMC and RNNs~\cite{Bennewitz2021NeuralEM, Czischek_2022}. Similarly, the incorporation of lattice symmetries provides a strategy to enhance the accuracy of our calculations~\cite{RNNWF, RNNAnnealing, Nomura2021}. Although our results match the anticipated behavior of the toric code and Bose-Hubbard spin liquid models, we highlight that the RNN wave function may be susceptible to spurious contributions to the TEE~\cite{kimUniversalLowerBound2023} and we have not addressed this issue in our work.

Finally, our methods can be applied to study other systems displaying topological order, such as the Rydberg atoms array~\cite{RubyDMRG2021, RydbergSimulator2021, giudici2022dynamical}, either through variational methods or in combination with experimental data. To experimentally study topological order, it is possible to use quantum state tomography with RNNs~\cite{Carrasquilla2019}. This involves using experimental data to reconstruct the state seen in the experiment followed by an estimation of the TEE using the methods outlined in our work. Overall, our findings suggest that RNN wave function ansatzes have promising potential for discovering phases of matter with topological order.


\section*{Acknowledgments}
We would like to thank Giacomo Torlai, Jeremy Côté, Schuyler Moss, Roeland Wiersema, Ejaaz Merali, Isaac De Vlugt, and Arun Paramekanti for helpful discussions. Our RNN implementation is based on Tensorflow~\cite{tensorflow2015-whitepaper} and NumPy~\cite{Harris2020}.
Computer simulations were made possible thanks to the Vector Institute computing cluster and Compute Canada. M.H acknowledges support from Mitacs through Mitacs Accelerate. We acknowledge support from Natural Sciences and Engineering Research Council of Canada (NSERC), the Shared Hierarchical Academic Research Computing Network (SHARCNET), Compute Canada, and the Canadian Institute for Advanced Research (CIFAR) AI chair program. Research at Perimeter Institute is supported in part by the Government of Canada through the Department of Innovation, Science and Economic Development and by the Province of Ontario through the Ministry of Colleges and Universities.
\appendix 

\section{2D periodic gated RNNs}
\label{app:GRU}
In this appendix, we describe our implementation of a 2D gated RNN wave function for periodic systems, that can be used to approximate the ground states of the Hamiltonians considered in this paper. If we define
\begin{align*}
    \bm{h'}_{i,j} &= [\bm{h}_{i-1,j} ; \bm{h}_{i,j-1}; \bm{h}_{\text{mod}(i+1,L_x),j}; \bm{h}_{i,\text{mod}(j+1, L_y)}], \\
    \bm{\sigma'}_{i,j} &= [\bm{\sigma}_{i-1,j} ; \bm{\sigma}_{i,j-1}; \bm{\sigma}_{\text{mod}(i+1,L_x),j}; \bm{\sigma}_{i,\text{mod}(j+1, L_y)}],
\end{align*}
then our gated 2D RNN wave function ansatz is based on the following recursion relations:
\begin{align*}
    \tilde{\bm{h}}_{i,j} &= \tanh \! \Big(
    W[\bm{\sigma'}_{i,j}; \bm{h'}_{i,j}]
    +  \bm{b} \Big), \\
   \bm{u}_{i,j} &= \text{sigmoid}\! \Big(
    W_g[\bm{\sigma'}_{i,j}; \bm{h'}_{i,j}]
    +  \bm{b}_g \Big), \\
    \bm{h}_{i,j} &= \bm{u}_{i,j} \odot \bm{\tilde{h}}_{i,j} + (1-\bm{u}_{i,j}) \odot (U \bm{h'}_{i,j}).
\end{align*}
Here `$\odot$' is the Hadamart (element-wise) product. A hidden state $\bm{h}_{i,j}$ can be obtained by combining a candidate state $\bm{\tilde{h}}_{i,j}$ and the neighboring hidden states $\bm{h}_{i-1,j}, \bm{h}_{i,j-1}, \bm{h}_{\text{mod}(i+1,L_x),j}, \bm{h}_{i,\text{mod}(j+1, L_y)}$. The update gate $\bm{u}_{i,j}$ determines how much of the candidate hidden state $\bm{\tilde{h}}_{i,j}$ will be taken into account and how much of the neighboring states will be considered. With this combination, it is possible to circumvent some limitations of the vanishing gradient problems~\cite{zhou2016minimal,shen2019mutual}. The weight matrices $W, W_g, U$ and the biases $b, b_g$ are variational parameters of our RNN ansatz. The size of the hidden state $\bm{h}_{i,j}$ is a hyperparameter called the number of memory units or the hidden dimension and is denoted as $d_h$. Finally, to motivate the use of the gating mechanism, we highlight that a gated 2D RNN was found to be better than a non-gated 2D RNN for the task of finding the ground state of a 2D Heisenberg model~\cite{RNNAnnealing}. 

Since we use an enlarged Hilbert space in our 2D RNN, we take $\bm{\sigma}_{ij}$ to be the concatenation of the one-hot encodings of the binary physical variables in each unit cell. In this case, if $m$ is the number of physical variables per unit cell, then $\bm{\sigma}_{ij}$ has size $2m$. We also note that the size of the Softmax layer is taken as $2^m$ so we can autoregressively sample each unit cell variables at once.

\section{Variational Monte Carlo and variance reduction}
\label{app:VMC}
To optimize the energy expectation value of our RNN wave function $\ket{\Psi_{\bm{\theta}}}$, we use the Variational Monte Carlo (VMC) scheme, which consists of using importance sampling to estimate the expectation value of the energy $E_{\bm{\theta}} = \bra{\Psi_{\bm{\theta}}} \hat{H} \ket{\Psi_{\bm{\theta}}}$ as follows~\cite{becca_sorella_2017,RNNWF}:
\begin{equation*}
    E_{\bm{\theta}} = \frac{1}{M}\sum_{i=1}^{M} E_{\text{loc}}(\bm{\sigma^{(i)}}),
\end{equation*}
where the local energies $E_{\text{loc}}$ are defined as
\begin{equation*}
    E_{\text{loc}}(\bm{\sigma}) = \sum_{\bm{\sigma'}} H_{\bm{\sigma} \bm{\sigma'}} \frac{\Psi_{\bm{\theta}}(\bm{\sigma'})}{\Psi_{\bm{\theta}}(\bm{\sigma})}.
\end{equation*}
Here the configurations $\{\bm{\sigma^{(i)}}\}_{i = 1}^{M}$ are sampled from our ansatz using autoregressive sampling. The choice of $M$ is a hyperparameter that can be tuned. Furthermore, $E_{\text{loc}}(\bm{\sigma})$ can be efficiently computed for local Hamiltonians. Furthermore, the gradients can be estimated as~\cite{RNNWF}
\begin{equation*}
    \partial_{\bm{\theta}} E_{\bm{\theta}} = \frac{2}{M}\mathfrak{Re} \left (\sum_{i=1}^{M} \partial_{\bm{\theta}} \log \left( \Psi_{\bm{\theta}}^{*}(\bm{\sigma^{(i)}}) \right) \left ( E_{\text{loc}}(\bm{\sigma^{(i)}}) - E_{\bm{\theta}} \right) \right).
\end{equation*}
For a stoquastic Hamiltonian $\hat{H}$~\cite{bravyi2015monte}, we can use a positive RNN wave function where the use of the real part $\mathfrak{Re}$ is not necessary. Importantly, subtracting the variational energy $E_{\bm{\theta}}$ is helpful to achieve convergence as it reduces the variance of the gradients near convergence without biasing its expectation value as shown in Ref.~\cite{RNNWF}. The subtracted term is referred to as a baseline, which is typically used for the same purpose in the context of Reinforcement learning~\cite{mohamed2019monte}. To demonstrate the noise reduction more rigorously compared to the intuition provided in Ref.~\cite{RNNWF}, let us focus on the variance of the gradient with respect to a parameter $\theta$ in the set of the variational parameters $\bm{\theta}$, after subtracting the baseline. Here we focus on the case of a positive ansatz wave function $\Psi_{\bm \theta}(\bm{\sigma}) = \sqrt{P_{\bm \theta}(\bm{\sigma})}$ that we used in our study. First of all, we define: 
\begin{equation*}
    O_{\theta}(\bm{\sigma}) \equiv \partial_{\theta} \log \left( \Psi^*_{\bm{\theta}}(\bm{\sigma}) \right) = \frac{1}{2} \partial_{\theta} \log( P_{\bm \theta}(\bm{\sigma})) .
\end{equation*}
Thus, the gradient with a baseline can be written as:
\begin{align*}
    \partial_{\theta} E_{\bm{\theta}} &= 2 \left \langle O_{\theta}(\bm{\sigma}) \overline{E}_{\text{loc}}(\bm{\sigma}) \right \rangle,
\end{align*}
where $\overline{E}_{\text{loc}}(\bm{\sigma}) \equiv E_{\text{loc}}(\bm{\sigma}) - E_{\bm{\theta}}$ and $\langle . \rangle$ denotes an expectation value over the Born distribution $|\Psi_{\bm{\theta}}(\bm{\sigma})|^2$. To estimate the gradients' noise, we look at the variance of the gradient estimator, which can be decomposed as follows:
\begin{align*}
    \text{Var} (O_{\theta} \overline{E}_{\text{loc}}) &= \text{Var}( O_{\theta} E_{\text{loc}} ) \\
    &- 2 \text{Cov}(O_{\theta} E_{\text{loc}} , O_{\theta} E_{\bm{\theta}}) + E_{\bm{\theta}}^2 \text{Var}(O_{\theta}).
\end{align*}
Thus the variance reduction $R$, after subtracting the baseline, is given as:
\begin{align*}
    R &\equiv \text{Var} (O_{\theta} \overline{E}_{\text{loc}}) - \text{Var}( O_{\theta} E_{\text{loc}} ), \\
    &= - 2 E_{\bm{\theta}} \text{Cov}(O_{\theta} E_{\text{loc}} , O_{\theta}) + E_{\bm{\theta}}^2 \text{Var}(O_{\theta}).
\end{align*}
Since the gradients' magnitude tends to near-zero values close to convergence, statistical errors are more likely to make the VMC optimization more challenging. We focus on this regime for this derivation to show the importance of the baseline in reducing noise. Thus, we assume that $E_{\text{loc}} (\bm{\sigma}) = E_{\bm{\theta}} + \xi(\bm{\sigma})$, where the supremum of the local energy fluctuations is much smaller compared to the variational energy, i.e., $(\sup_{\bm{\sigma}} |\xi(\bm{\sigma})|) \ll E_{\bm{\theta}}$. From this assumption, we can deduce that:
\begin{align}
    R &= - 2 E_{\bm{\theta}}^2 \text{Cov}(O_{\theta} , O_{\theta}) \\
    &- 2 E_{\bm{\theta}} \text{Cov}(O_{\theta} \xi , O_{\theta}) 
     + E_{\bm{\theta}}^2 \text{Var}(O_{\theta}), \\
    &= - E_{\bm{\theta}}^2 \text{Var}(O_{\theta}) - 2 E_{\bm{\theta}} \text{Cov}(O_{\theta} \xi , O_{\theta}).
    \label{eq:variance_reduction}
\end{align}
The second term can be decomposed as follows:
\begin{align}    \text{Cov}(O_{\theta} \xi , O_{\theta}) &= \langle O_{\theta}^2 \xi \rangle - \langle O_{\theta}\xi\rangle \langle O_{\theta} \rangle .
\label{eq:cov_decomp}
\end{align}
Since $\langle O_{\theta} \rangle =  \frac{1}{2}\langle \partial_{\theta} \log( P_{\bm \theta}) \rangle = 0$~\cite{sutton2000policy, mohamed2019monte,RNNWF}, then we can bound the covariance term from above as:
\begin{align*} \text{Cov}(O_{\theta} \xi , O_{\theta}) &\leq  \left(\sup_{\bm{\sigma}} |\xi(\bm{\sigma})| \right) \langle O_{\theta}^2 \rangle, \\
&= \left(\sup_{\bm{\sigma}} |\xi(\bm{\sigma})| \right) \text{Var}(O_{\theta}), \\
    &\ll E_{\bm{\theta}} \text{Var}(O_{\theta}).
\end{align*}
Thus, we can conclude that the variance reduction $R$ in Eq.~\eqref{eq:variance_reduction} is negative. This observation highlights the importance of the baseline in reducing the statistical noise of the energy gradients near convergence. 

For a complex ansatz wave function $\Psi_{\bm \theta}(\bm{\sigma}) = \sqrt{P_{\bm \theta}(\bm{\sigma})} \exp\left(\text{i} \phi_{\bm \theta} (\bm{\sigma})\right)$, the expectation value $\langle O_{\theta} \rangle =  \frac{1}{2}\langle \partial_{\theta}\log( P_{\bm \theta}) \rangle - \text{i} \langle \partial_{\theta} \phi_{\bm \theta} \rangle = - \text{i} \langle \partial_{\theta} \phi_{\bm \theta}\rangle$ is no longer equal zero in general. We leave the investigation of this case for future studies.

Similarly to the stochastic estimation of the variational energy using our ansatz, we can do the same for the estimation of the variational pseudo free energy $F_{\bm \theta}$ in Eq.~\eqref{eq:FreeEnergy}. More details can be found in the supplementary information of Ref.~\cite{VNA2021}. Finally, we note that the gradient steps in our numerical simulations are performed using Adam optimizer~\cite{AdamPaper}.

\section{Hyperparameters}
\label{app:hyperparams}

For all models studied in this paper, we note that for each annealing step, we perform $N_{\rm train} = 5$ gradient steps. Concerning the learning rate $\eta$, we choose $\eta = 10^{-3}$ during the warmup phase and the annealing phase and we switch to a learning rate $\eta = 10^{-4}$ in the convergence phase. We finally note that we set the number of convergence steps as $N_{\rm convergence} = 10000$. In Tab.~\ref{tab:hyperparams}, we provide further details about the hyperparameters we choose in our study for the different models. The meaning of each hyperparameter related to annealing is discussed in detail in Refs.~\cite{VNA2021, RNNAnnealing}.

Additionally, we use $M_e = 2 \times 10^6$ samples for the estimation of the entanglement entropy along with their error bars for the toric code. For the Bose-Hubbard model we use $M_e = 10^7$ samples to reduce the error bars on the TEE in Fig.~\ref{fig:BoseHubbard}. To estimate the TEE uncertainty from the Kitaev-Preskill construction, we use the standard deviation expression of the sum of independent random variables~\cite{ku1966notes}.

Finally, we note that to avoid fine-tuning the learning rate for each value of $V$ (between $4$ and $13$) in the Bose-Hubbard model, we target the normalized Hamiltonian
\begin{equation}
    \hat{H} = -\frac{1}{V} \sum_{\langle i,j \rangle} \left( b_i^{\dagger} b_j + b_i b_j^{\dagger} \right) + \sum_{\hexagon} n_{\hexagon}^2
\end{equation}
in our numerical experiments.

\begin{table*}[htp]
    \centering
    \scriptsize
    \begin{tabular}{|c|c|c|}\hline
       Figures & Parameter & Value \\\hline
      \multirow{4}{*}{2D toric code} & Number of memory units & $d_h = 60$ \\
            & Number of samples & $M = 100$ \\
            & Initial pseudo-temperature & $T_0 = 2$ \\
            & Number of annealing steps
             & $N_{\rm annealing} = 4000$ \\             
       \hline
      \multirow{4}{*}{Bose-Hubbard model (L = 6)} & Number of memory units & $d_h = 100$ \\
            & Number of samples & $M = 500$ \\
            & Initial pseudo-temperature & $T_0 = 1$ \\
            & Number of annealing steps
             & $N_{\rm annealing} = 10000$ \\            
       \hline     
         \multirow{4}{*}{Bose-Hubbard model (L = 8)} & Number of memory units & $d_h = 100$ \\
        & Number of samples & $M = 500$ \\
        &  Pseudo-temperature & $T_0 = 0$ \\
        & Number of steps
         &  $10000$ \\             
       \hline     
          
    \end{tabular}
    \caption{A summary of the hyperparameters used to obtain the results reported in this paper. Note that the number of samples $M$ corresponds to the batch size used during the training phase. Additionally, the $L = 8$ RNN model is initialized with the $L=6$ optimized RNN for the Bose-Hubbard model.} 
    \label{tab:hyperparams}
\end{table*}



\clearpage
\bibliography{Biblio}

\end{document}